\pdfoutput=1
\documentclass[a4paper, 10pt]{scrartcl}
\usepackage[english]{babel}
\usepackage{graphicx}
\usepackage[utf8]{inputenc}
\usepackage{fixltx2e}
\usepackage{url}
\usepackage{amsmath,amssymb}
\usepackage{authblk}
\usepackage[small]{caption}
\usepackage[colorlinks=true, citecolor=blue, linkcolor=blue, filecolor=blue,
urlcolor=blue]{hyperref}
 \newcommand{\sto}{\operatorname{subject\ to:\ }}
 \newcommand{\figdir}{.}
\begin{document}

\title{\vspace{-3ex}What can transmission do for a fully
        renewable Europe?}
\author[1]{\small Sarah Becker\thanks{Corresponding author, e-mail:
\href{mailto:becker@fias.uni-frankfurt.de}{becker@fias.uni-frankfurt.de}}}
\author[2]{\small Rolando A. Rodriguez}
\author[2]{\small Gorm B. Andresen}
\author[2]{\small Martin O.W. Greiner}
\author[1]{\small Stefan Schramm}

\affil[1]{\small FIAS Frankfurt, Frankfurt University, Germany}
\affil[2]{\small Department of Engineering and Department of Mathematics, Aarhus University, Denmark}

\date{\small September 24, 2013}

%
%
%

\maketitle

\begin{abstract}
  {\small {\sc Abstract.} Our research is centred around the question how to
    best integrate the variable renewable energy sources (VRES), wind power and
    solar photovoltaics, into the European electricity grid. The future
    electricity supply will be based to a large extend on these fluctuating
    resources. We have conducted a study, extrapolating national historical and
    targeted wind and solar power penetrations in Europe up to 100\% VRES
    \cite{rolando,sarah}. A high share of VRES means large fluctuations in the
    generation, causing overproduction and deficits.  One way to reduce such
    mismatches is power transmission spatially smoothing out the fluctuations.
    This has the potential to reduce the remaining shortages by sharing the
    surplus production of others. We find that shortages can at maximum be
    reduced by 40\% in the hypothetical case of unlimited transmission
    capacities across all of Europe. A more realistic extension of the
    transmission grid, roughly quadrupling today's installation, turns out to be
    sufficient to harvest 90\% of this potential benefit. Finally, the import
    and export of single countries is investigated. We conclude that a country's
    load size as well as its position in the network are the determining factors
    for its import/export opportunities.
  }
\end{abstract}

\section{Motivation}

The ambitious goal of the European Union to reduce green house gases, in
particular CO$_2$ by 80\% by 2050 \cite{eu2050}, entails a shift to almost
completely CO$_2$-free electricity generation in the same time horizon
\cite{ecf2050}. A large part of this renewable generation will come from the
variable renewable energy sources (VRES) wind and solar PV. Due to their
intermittent nature, their integration into the power system poses a
considerable challenge. To this end, various measures have been proposed and to
some extend also tested or implemented: Energy storage systems, demand-side
management (DSM)/Smart Grid technologies, or coupling to the other energy
sectors, heating/cooling and transportation (via electrical heating/cooling,
electric vehicles, or fuel generation from electricity and vice versa), all have
the potential to shift load and/or generation in time, thus reducing the
mismatch between the two. Another path to follow to reduce the mismatch between
load and generation is its spatial distribution over larger areas, i.e.\ power
transmission. Wind power production, for example, has been shown to decorrelate
over the range of $\sim$1000~km for the cases of the US East Coast
\cite{Kempton:2010oq} and Sweden \cite{Widen:2011ys}. Such spatial smoothing has
the potential to yield a more even generation output, and hence a better chance
to match it to the load. In this proceedings, we assess the potential of
transmission from a purely technical point of view. We investigate what
transmission is able to do as well as what it cannot do.  Economics are not
taken into account. Furthermore, we look for efficient ways of enhancing the
transmission grid and develop a tentative roadmap of which links should be
reinforced and when, and  we explore the effects of such an enhanced
transmission grid on the trade opportunities of single countries. Parts of these
results have been published in \cite{rolando,sarah}. Here, we present additional
discussion of the factors that influence a country's import and export
opportunities, and show the time development of the distributions of the
mismatch between load and renewable generation. The latter reveal that
transmission is particularly well-suited to reduce the occurence of small
mismatch events, but the mitigation of large mismatches remains a major
challenge, even when an upgraded transmission grid is available.

\section{Data and methodology}

\subsection{Weather and generation}

In this paper, we use the same weather and generation data that have been
described at this conference in the talk "Weather-driven modelling of future
renewable power systems" by G.\ B.\ Andresen et.\,al. Therefore, we only repeat
the essentials of our Weather-Driven Renewable Energy System Modelling (WDRESM).
At the core lie generation data for wind and solar PV power which are calculated
based on real weather data of eight years with hourly resolution. They are
aggregated to country level and normalized by their mean, yielding $W_n(t)$
(wind time series for country $n$) and $S_n(t)$ (solar time series for country
$n$). Load data $L_n(t)$ were obtained for the same eight years, in hourly
resolution as well.  The key ingredient to all further modelling is the mismatch
between generation and load:
\begin{align}
    \Delta_n(t) = \gamma_n \left(\alpha_n^W W_n(t) + (1-\alpha_n^W)
    S_n(t)\right) \cdot \langle L_n \rangle - L_n(t)
    \label{eq:mism}
\end{align}
In this formula, the factor $\gamma_n$ scales the average generation from the
two variable renewable sources (VRES) to a desired fraction of the load, so we
can model different penetrations of wind and solar power, and $\alpha_n^W$ is
the relative share in wind in VRES. $\langle .\rangle$ denotes the time average
of a quantity. If this mismatch is positive, it means that surplus generation
occurs which can be exported to other countries or used domestically in other
energy sectors. If it is negative, generation is insufficient and the electrical
load has to be covered from other sources or shifted to other times.

\subsection{Logistic fit}
\label{sec:logfit}

\begin{figure}[!ht]
    \begin{center}
        \includegraphics[width=0.49\textwidth]{\figdir/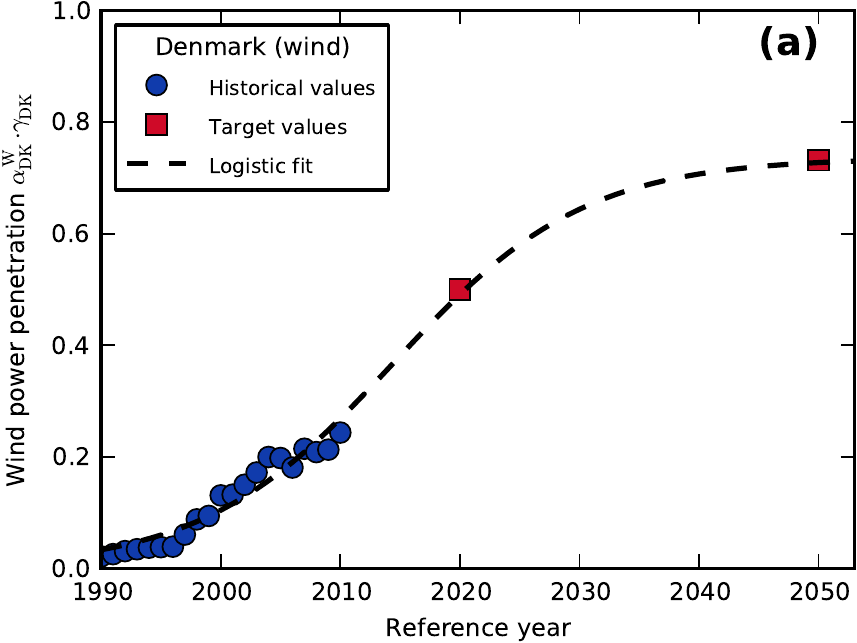}
        \includegraphics[width=0.49\textwidth]{\figdir/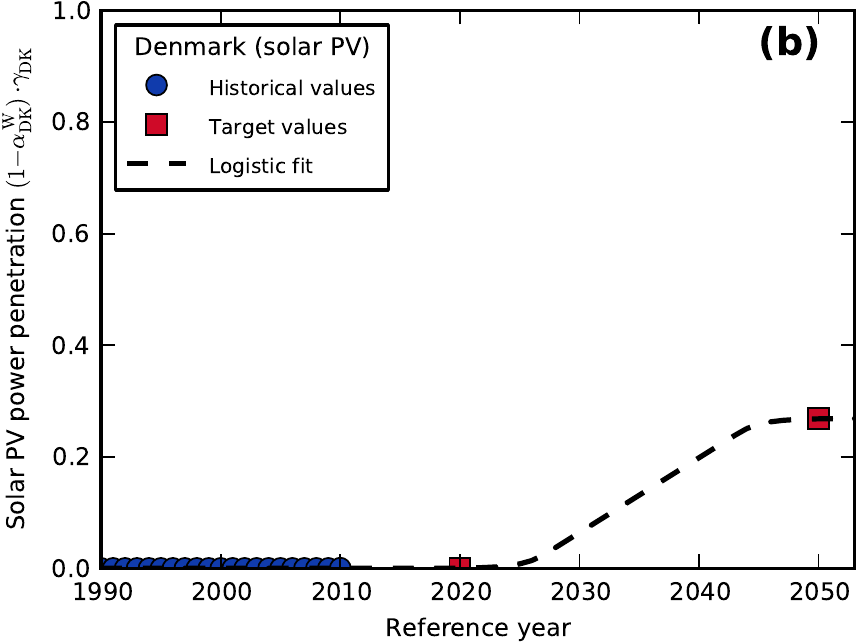}
        \caption{Logistic fit of wind power penetration (a) and solar power
                penetration (b) for the example case of Denmark. The
                historical values originate from eurostat \cite{eurostat}, the
                2020 targets from the National Renewable Energy Action Plan
                \cite{nreap}, and the 2050 target is a hypothetical scenario in
                which the average wind and solar power generation equals the
                average load.
        }
        \label{fig:logfit}
    \end{center}
\end{figure}

In order to have a tangible scaling of the renewable penetration from the
medium-sized percentage seen today to a scenario where mean load equals mean
VRES generation, we extrapolate logistically from historical values through the
2020 EU targets \cite{nreap} to a hypothetical 2050 target of an on average
100\% VRES-supplied European electricity system. This still includes balancing
from dispatchable sources. For details of the data as well as the logistic fit,
see \cite{sarah}. An example is shown in Fig.\ \ref{fig:logfit}.

\subsection{Power transmission between countries}
\label{sec:flows}

In our power transmission model, each country forms a node in a network. In the
case at hand, we investigate the European countries. The topology of the network
can be seen in Fig.\ \ref{fig:ntx}. The power transmission
framework has been developed and tested as published in \cite{rolando,sarah}. It
consists of a generalized DC power flow. For a stable electricity grid, the full
AC equations can be well approximated linearly, yielding a system that is
mathematically completely analogous to DC power flow \cite{ooe}. One formulation
of this is the following: Given a mismatch vector, $(\Delta_n)_{n=1..N}$,
find a flow vector $(F_l)_{l=1..L}$ along the $L$ links of the network that solves
\begin{align}
    \Delta_n -(K F)_n = 0 & \qquad \text{(cover deficits with surpluses)}\\
    \min \sum_l F_l^2 & \qquad \text{(minimize transmission dissipation)},
\end{align}
where $F_l$ is the power flow along link $l$ and $K$ is the incidence matrix, 
\begin{align*}
    K_{nl} = \begin{cases}
            \phantom{-}1 & \text{ if link $l$ starts at node $n$} \\
            -1  & \text{ if link $l$ ends at node $n$} \\
            \phantom{-}0 & \text{ else}
    \end{cases}.
\end{align*}
This has two drawbacks that make it seem unattractive for our application: It
only has a solution if $\sum_n\Delta_n = 0$, and it does not allow to take
limits on the flow due to finite line capacities into account. We therefore
generalize it in two ways: First, we allow for residual surpluses and deficits.
The latter have to be covered from other sources. We call those "balancing",
since they balance the system, $B_n(t)$. We minimize their usage such that VRES
are always preferred. Second, we include the constraints on the flows in the two
resulting minimizations:
\begin{align}
    \label{eq:step1}
    \begin{split}
        \text{1st step: } & \min\sum_n(\Delta_n -(KF)_n)_- = \min\sum_n B_n
        \rightarrow
        B_\text{min} \quad \text{(minimize back-up power)}\\
        &\sto h_{-l} \leq F_l \leq h_l
    \end{split}
\end{align}
\begin{align}
    \label{eq:step2}
    \begin{split}
      \text{2nd step: } & \min \sum_l F_l^2 
      \qquad \qquad \qquad \qquad \qquad \qquad \qquad \quad \quad \ \ 
      \text{(minimize dissipation)}
      \qquad \\
        &\sto h_{-l} \leq F_l \leq h_l,\quad
        \sum_n B_n = B_\text{min}
    \end{split}
\end{align}
Here, $(x)_- = \max\{0,-x\}$ denotes the negative part of a quantity $x$. This
optimization problem is proven to be convex and can hence be handled with
standard techniques.

As a last remark, we observe that this flow paradigm leads to power flow only
transporting VRES surplus production from one country to another, where it
either covers VRES deficits or replaces balancing. Trade with conventional
power, balancing in our terminology, does not take place. This follows from Eq.\
\ref{eq:step2}: Producing the balancing energy for one country in another and
then transferring it leads to more flow while keeping the total amount of
balancing constant.  Minimizing the flow means localising the balancing at those
nodes that actually see a deficit.

\section{Transmission}

\subsection{Benefit of transmission}

The benefit of transmission of a certain line capacity layout is formulated as a
percentage of reduction in balancing energy relative to what is achievable with
transmission. Denote a given line capacity layout, i.e.\ a set of line
capacities, by $CL_\text{this}=\{h_{\pm l}^\text{(this)}\}_{l=1..L}$.  Under the
fixed flow paradigm defined above, this gives rise to a well-defined
corresponding total balancing energy needed, $B_\text{tot}(CL_\text{this}) =
\sum_n \sum_t B_n(t)$. For unconstrained transmission,
$CL_\text{unconstrained}$, as well as for no transmission, $CL_\text{zero}$, the
total balancing energy is calculated.  The difference between those two is the
maximal balancing energy reduction achievable by transmission.  The benefit of
$CL_\text{this}$ is now defined as the balancing energy reduction resulting from
going from $CL_\text{zero}$ to $CL_\text{this}$, normalized by the maximal
possible reduction.  
\begin{align}
    \beta_\text{this} = \frac{B_\text{tot}(CL_\text{zero}) -
    B_\text{tot}(CL_\text{this})}{B_\text{tot}(CL_\text{zero}) - B_\text{tot}(CL_\text{unconstrained})}
\end{align}
Fixing the benefit of transmission amounts to fixing the total balancing energy.

\subsection{Build-up of the transmission grid}
\label{sec:bu}

\begin{figure}[!b]
    \begin{center}
        \includegraphics[width=0.40\textwidth]{\figdir/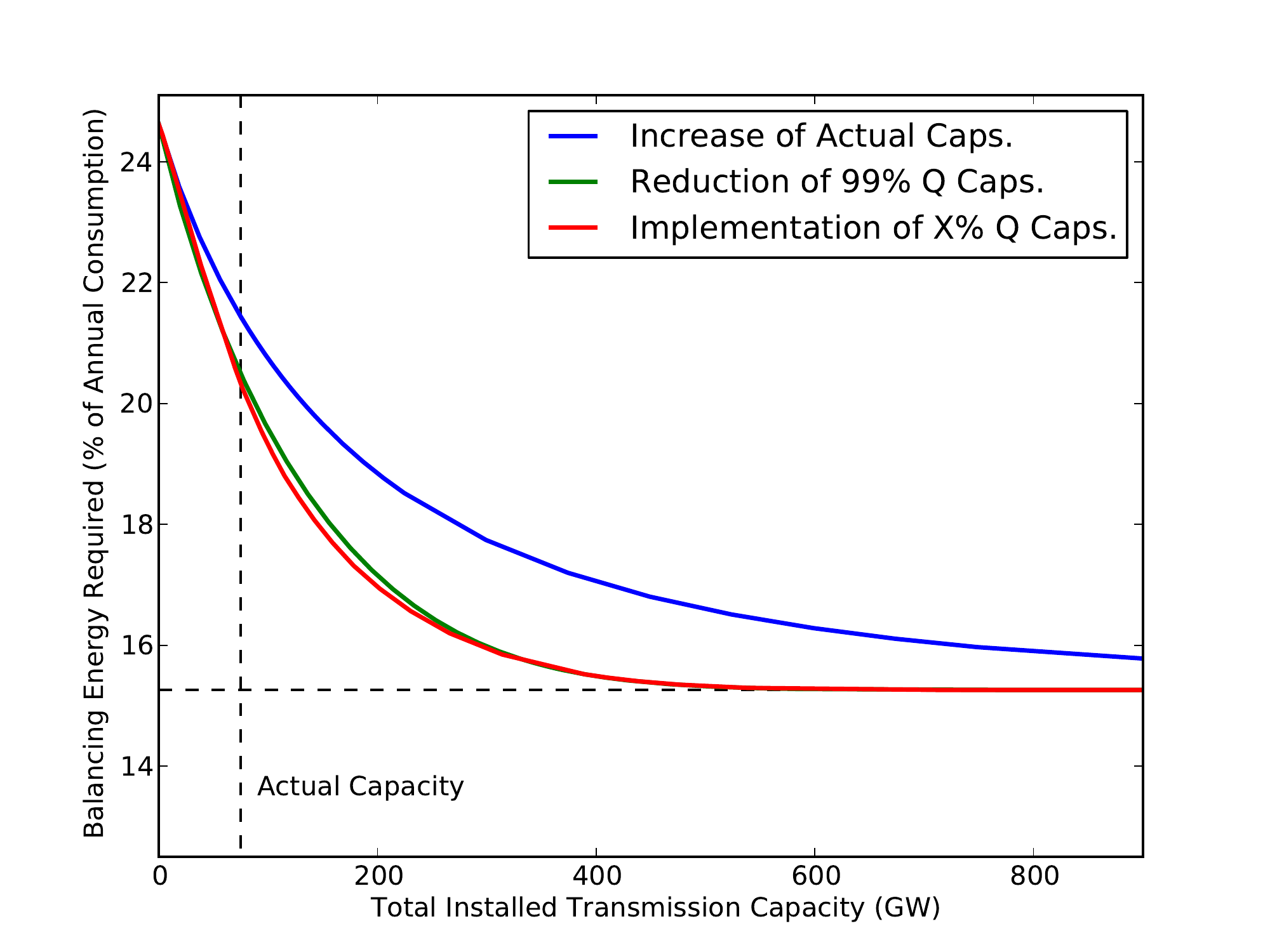}
        \includegraphics[width=0.55\textwidth]{\figdir/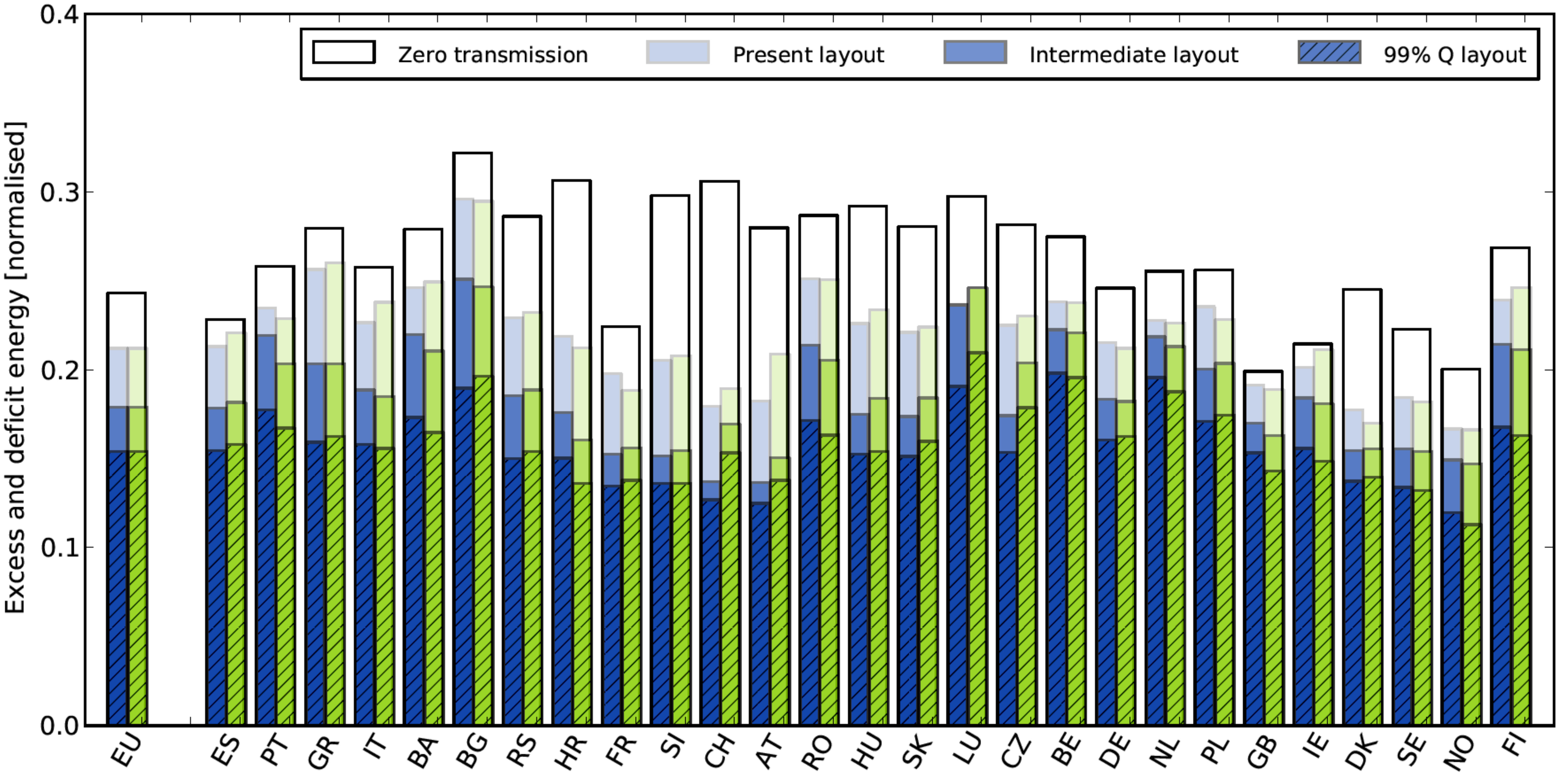}
        \caption{Left: Normalized total balancing energy as a function of
            transmission strength. The vertical dashed line indicates the total
            installed capacity in winter 2010/2011, as reported by
            \cite{entsoe_ntc}. The horizontal dashed line is the asymptotic
            limit for strong transmission grids. Notice how the balancing energy
            can be reduced by 40\% by transmission. The different curves
            correspond to different distributions of capacity across the single
            lines, see Sec.\ \ref{sec:bu} for details. Right: Balancing energy
            ("deficit", shown in blue) and surplus ("excess", shown in green)
            energy for single countries, for different transmission layouts:
            Zero transmission, today's (winter 2010/2011) line capacities, 99\%
            quantiles of the unconstrained flow, and an intermediate layout
            halfway between today's and the 99\% quantile layout. Right panel
            taken from \cite{rolando}. 
        }
        \label{fig:BT}
    \end{center}
\end{figure}

An obvious question to ask, given a transmission grid topology, is how to extend
it in the most efficient way. If we had a certain investment (in terms of
additional MW in line capacity), how should it be distributed among the links to
obtain the highest possible benefit of transmission? We have tested several
build-up scenarios, as described in \cite{rolando}, for the case of
$\gamma_n=1\,\forall\,n$ in Eq.  \eqref{eq:mism}, i.e.\ VRES generation average
equalling load average, and a wind/solar mix $\alpha_n^W$ that makes their
generation follow the load in the closest possible way, for details on this, see
\cite{rolando}. The first (blue curve in Fig.\ \ref{fig:BT}a) is an upscaling of
the line capacities present today.
\begin{align}
    h_l^{(1)} = b\cdot h_l^\text{today}\,\forall \, l
\end{align}
For the second (red curve in Fig.\ \ref{fig:BT}a), we calculate the flows
resulting from removing the flow constraint from the two minimizations
\eqref{eq:step1}, \eqref{eq:step2}. A capacity layout is then constructed by
taking quantiles $Q$ of the distributions of these unconstrained
flows:
\begin{align}
    h_l^{(2)} = \operatorname{Quantile}^Q(\{F_l(t)\}_{t=1..T})\,\forall \, l
\end{align}
The third (green curve in Fig.\ \ref{fig:BT}a) is a simplification of the
second: Instead of taking different quantiles, only the 99\% quantiles are used
and scaled down. This yields similar capacities as compared to the second
approach.
\begin{align}
    h_l^{(3)} = b\cdot\operatorname{Quantile}^{99\%}(\{F_l(t)\}_{t=1..T})\,\forall \, l
\end{align}

The most successful interpolation scheme of the three under consideration are the
quantile capacities, the second method described above, as can be seen in Fig.\
\ref{fig:BT}a. This will be used as a standard for the rest of this proceedings.

When looking at the performance of the different transmission enhancement
schemes (Fig.\ \ref{fig:BT}), we see that transmission can, at best, reduce the
residual need for balancing by about 40\%, from 24\% of the load to 15\% of the
load, but not more. The rest has to be dealt with by other measures, such as
DSM or coupling to other energy sectors, no matter how strong the transmission
grid.

\begin{figure}[p]
    \begin{center}
        \includegraphics[width=0.32\textwidth]{\figdir/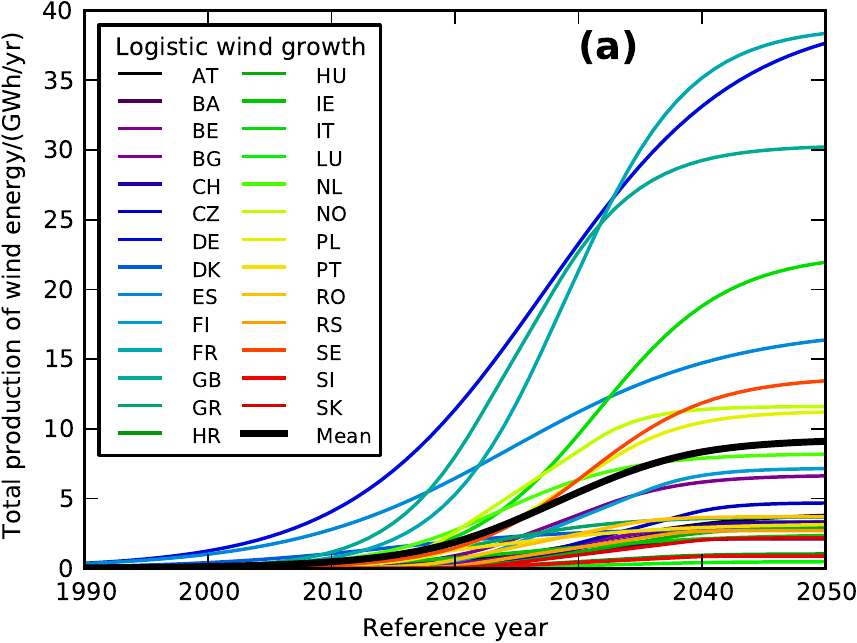}
        \includegraphics[width=0.32\textwidth]{\figdir/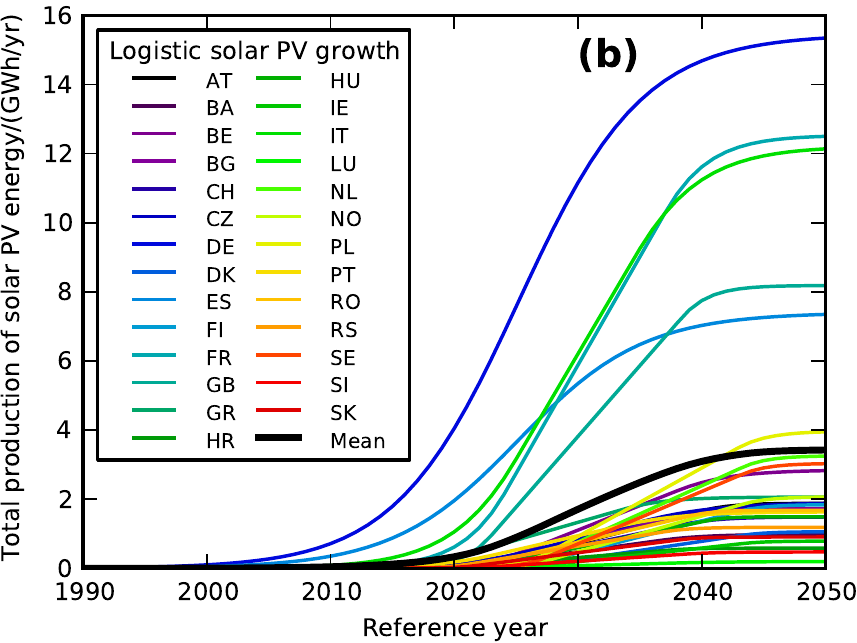}
        \includegraphics[width=0.32\textwidth,type=pdf,ext=.pdf,read=.pdf]{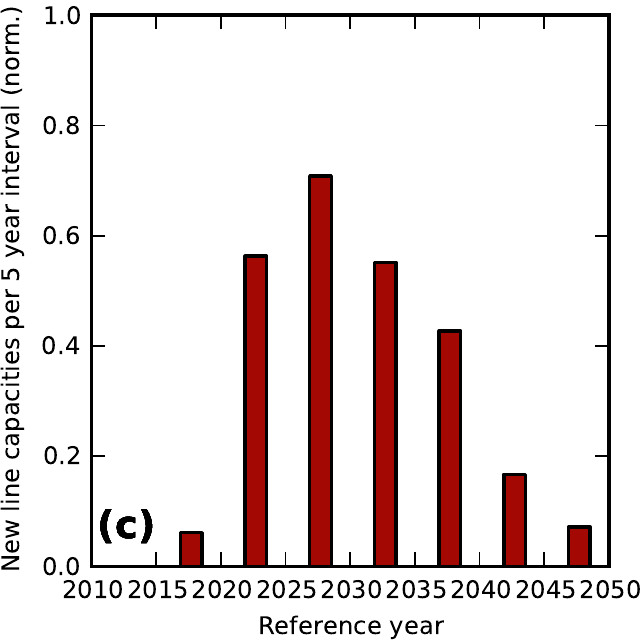}
        \caption{Growth of wind as well as solar installations for the single
            countries for wind (a) and solar PV (b), and the line investment per
            five-year interval necessary to keep the line capacities at a level
            of 90\% benefit of transmission, panel (c).
        }
        \label{fig:growth}
    \end{center}
\end{figure}
\begin{figure}[p]
    \begin{center}
        \includegraphics[width=0.49\textwidth]{\figdir/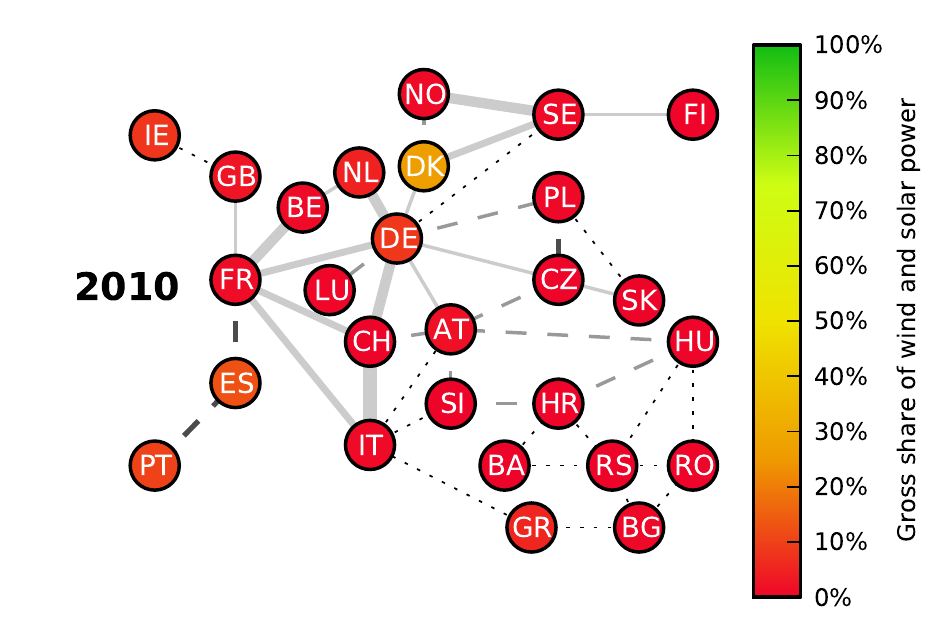}
        \includegraphics[width=0.49\textwidth]{\figdir/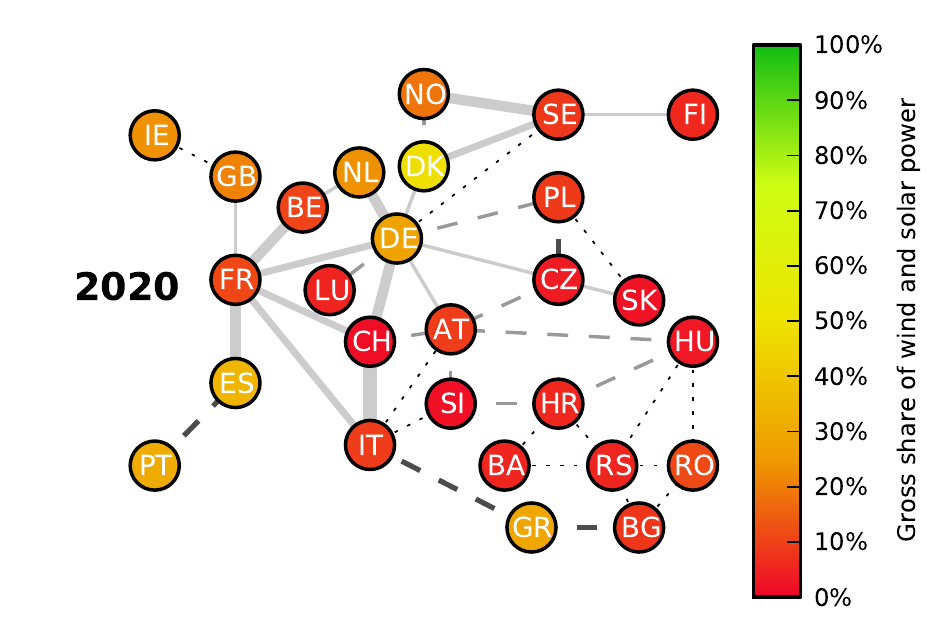}
        \includegraphics[width=0.49\textwidth]{\figdir/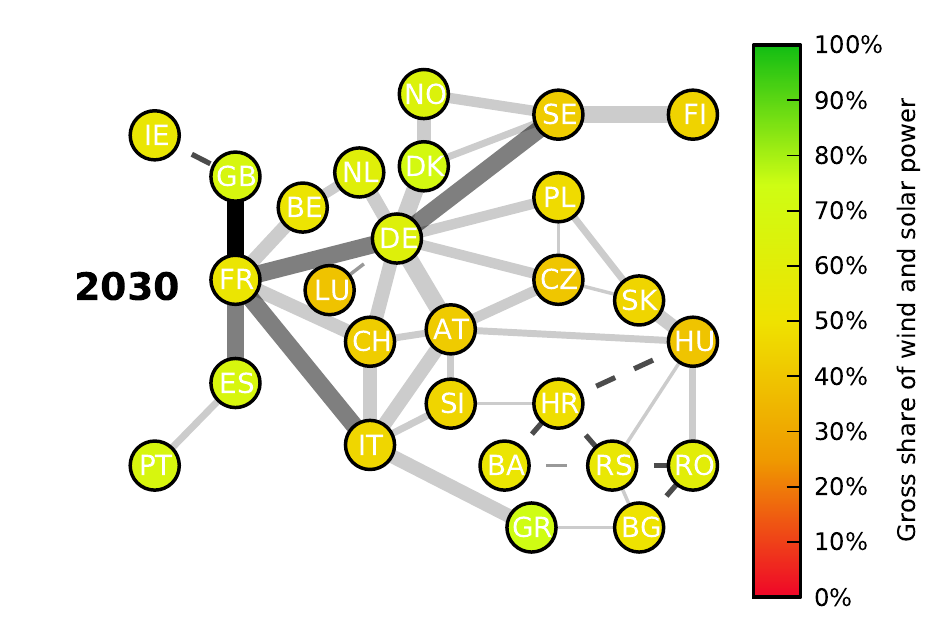}
        \includegraphics[width=0.49\textwidth]{\figdir/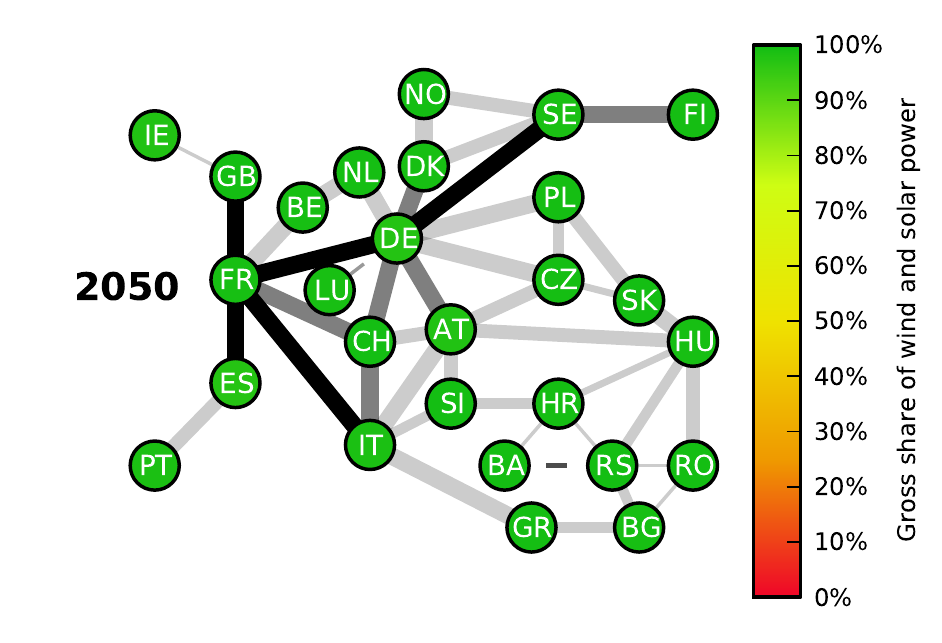}
        \includegraphics[width=0.49\textwidth]{\figdir/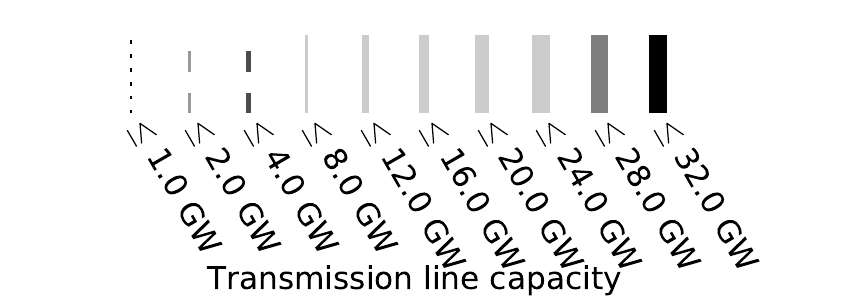}
        \caption{Growth of the European transmission network necessary to
            provide a 90\% benefit of transmission throughout the years. Shown
            are snapshots of 2010, 2020, 2030, and 2050. Line style and
            thickness indicates its transmission capacity, but node sizes and
            line lengths are not to scale. The VRES gross share $\gamma_n$ is
            colour-coded for each of the nodes $n$, from red ($\gamma_n=0.0$)
            through yellow ($\gamma_n=0.5$) to green ($\gamma_n=1.0$).
        }
        \label{fig:ntx}
    \end{center}
\end{figure}

\subsection{Line build-up in time}

In the previous section, a scenario has been examined where VRES installations
provide already (on average) as much as what is consumed, while the lines are
build up. To make the model more realistic, we now turn to a scenario where the
lines are build in parallel to the VRES installations. The VRES build-up is
assumed to follow the logistic growth discussed in Sec.\ \ref{sec:logfit}. The
line capacities are chosen as quantiles of the unconstrained capacities of the
fully renewable 2050 scenario, such that a 90\% benefit of transmission is
maintained throughout the years.  In our model, the build-up in wind
installations takes place mainly between 2015 and 2035, and solar PV
installations follow with a delay of about five years, see Fig.\
\ref{fig:growth}a and b.  Correspondingly, the line build-up necessary to keep
the benefit of transmission at 90\% is highest between 2020 and 2035, see Fig.\
\ref{fig:growth}c. The line capacities are summed over all links and normalized
by the total installations we have today. Overall, they have to be quadrupled as
compared to today's values over the course of the years. The development of the
single lines as well as the nodes is shown in Fig.\ \ref{fig:ntx}. It is seen
that in particular, the weak link between Spain and France as well as between
Great Britain and continental Europe are dramatically reinforced, while the grid
in the South-East remains relatively weak.

\subsection{Import and Export}

\begin{figure}[!b]
  \begin{center}
    \includegraphics[width=0.32\textwidth]{\figdir/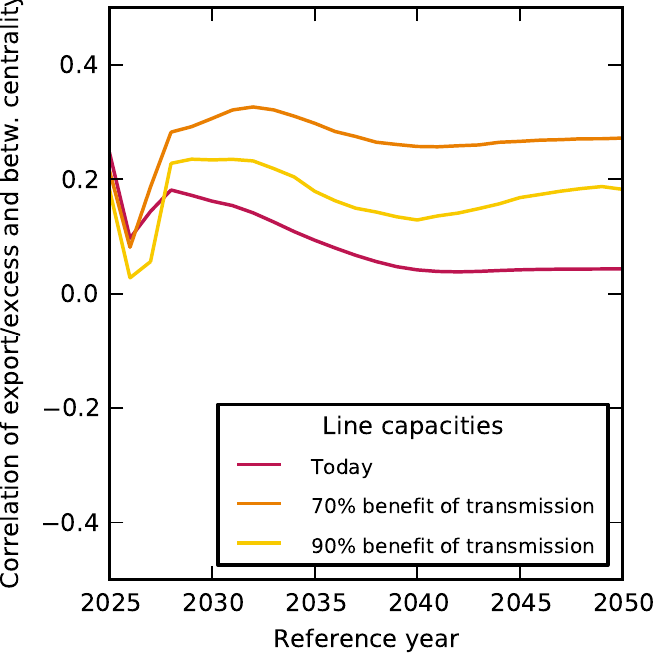}
    \includegraphics[width=0.32\textwidth]{\figdir/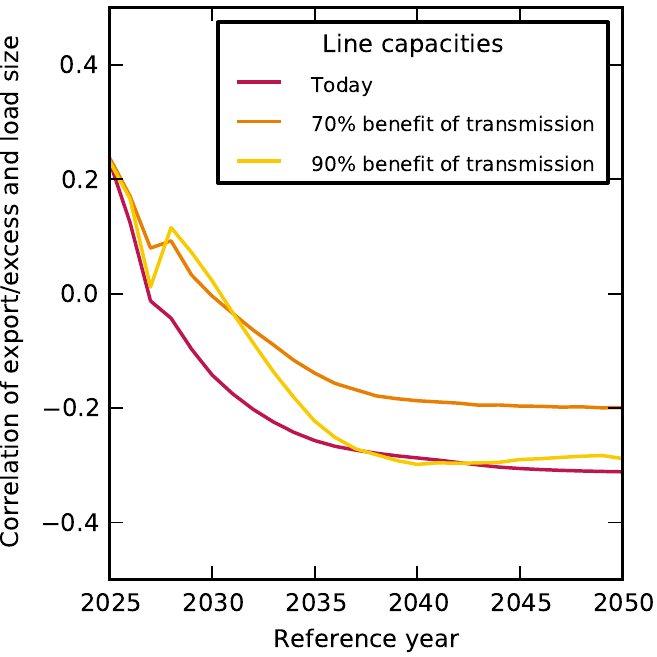}
    \includegraphics[width=0.32\textwidth]{\figdir/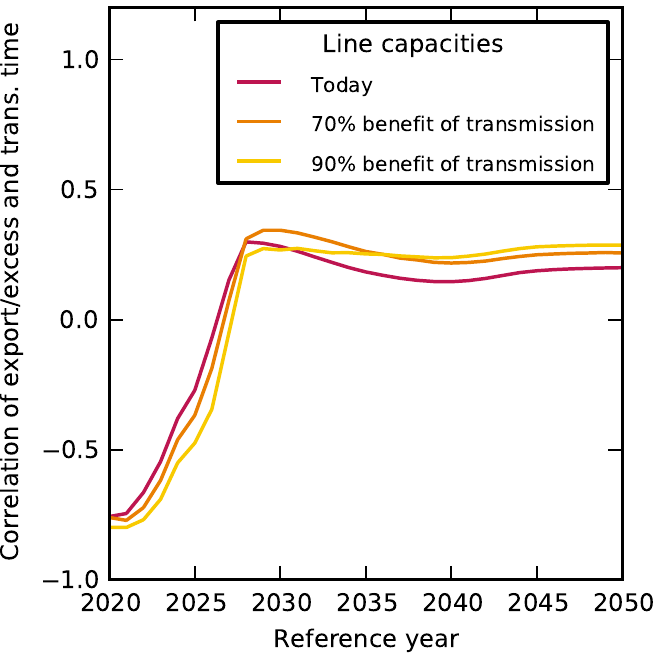}
    \caption{Correlation across all countries between exported surplus fraction
      and (a) betweenness centrality, a widely used measure from the theory of
      complex networks of how central a node is, (b) load size, and (c)
      transition time.
    }
    \label{fig:cor}
  \end{center}
\end{figure}

From the single countries' perspectives, it is also interesting to see how they
fare in the international power trade. Keep in mind that we only use power flow
to distribute VRES excess from one country to the other; conventional generation
is not transported (Sec.\ \ref{sec:flows}). As is already seen in Fig.\
\ref{fig:BT}b, the different countries do not have equal import and export
opportunities. Some can reduce their relative surplus as well as their relative
deficits more than others. We find that there are in general three factors that
have an influence on the fraction of surplus that can be covered by
imports/exports: Firstly, size matters. While the relative deficits of e.g.\
Germany and Denmark are comparable, the absolute values are not. This means that
big Germany has far worse chances of covering its deficit by imports, while
small Denmark encounters relatively few problems. The same holds mutatis
mutandis for exports. The second factor is the time of transition. As a proxy,
we use here the year in which the gross penetration $\gamma$ reaches 50\%. Early
adopters, such as Denmark or Spain, face an export boom in the beginning when
they see surplus production while others do not yet. Since our power
distribution favours VRES whenever possible, this surplus production can almost
certainly be exported to other countries where it replaces balancing. The third
factor is the position of the country in the network, whether it is central or
peripheral. This is due to the flow minimization we perform, see Eq.\
\eqref{eq:step2}.  Since flow to or from a central country comes with shorter
paths, on average, it is preferred. Therefore, central countries have slightly
better import and export opportunities than peripheral ones. The correlations
between the exported fraction of the surplus and these three factors are shown
in Fig.\ \ref{fig:cor}.
\begin{figure}[p]
  \begin{center}
    \includegraphics[width=0.89\textwidth]{\figdir/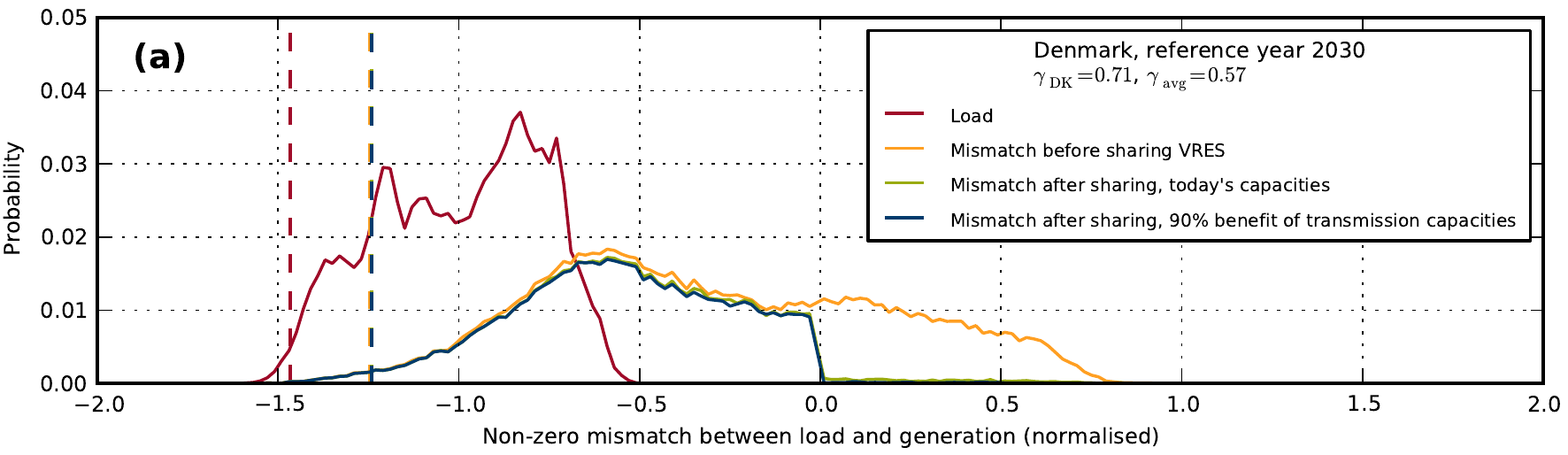}
    \includegraphics[width=0.89\textwidth]{\figdir/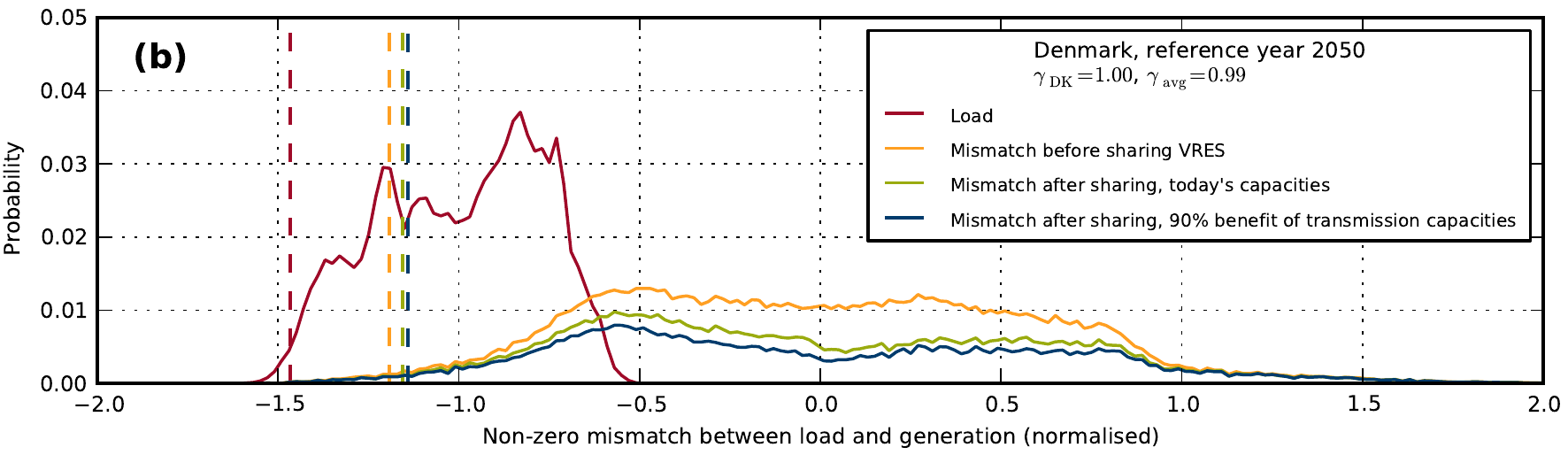}
    \caption{Distribution of the load (red), distribution of the mismatch
      between renewable generation and load before power sharing takes place
      (yellow), after sharing takes place with today's line capacities
      (green), and after sharing takes place with the 90\% benefit of
      transmission line capacities (blue), for the years 2030 (a) and 2050
      (b), for Denmark. The dashed lines indicate the 99\% quantile of the
      residual deficit. For clarity, the peak at zero is not shown here.
      Some of the lines cover each other.
    }
    \label{fig:mmDK}
  \end{center}
\end{figure}
\begin{figure}[p]
  \begin{center}
    \includegraphics[width=0.89\textwidth]{\figdir/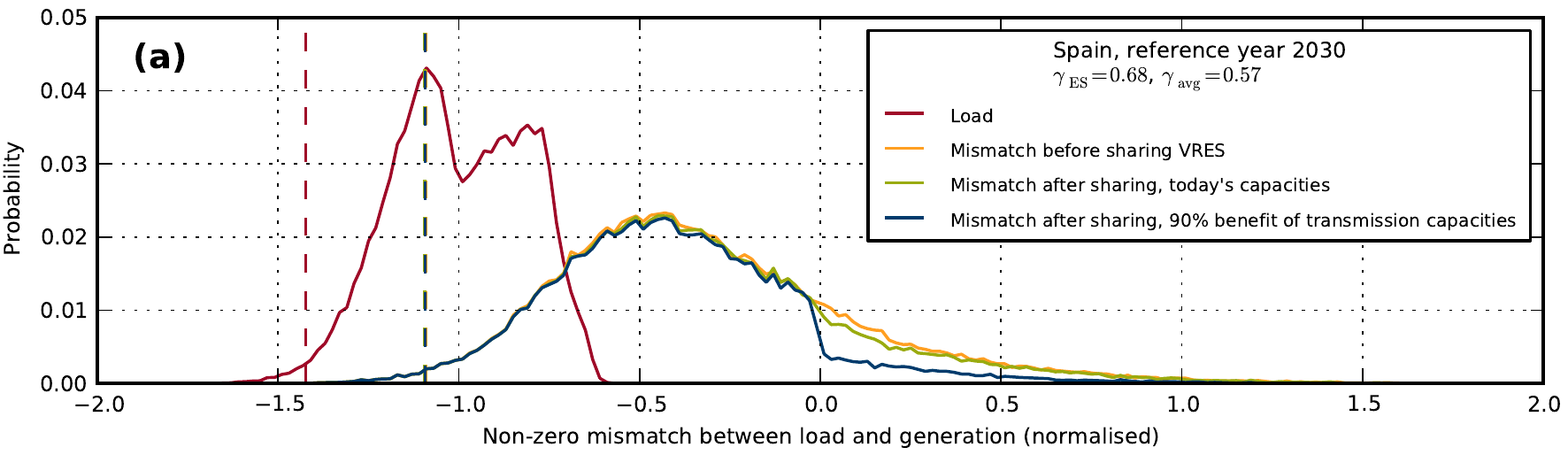}
    \includegraphics[width=0.89\textwidth]{\figdir/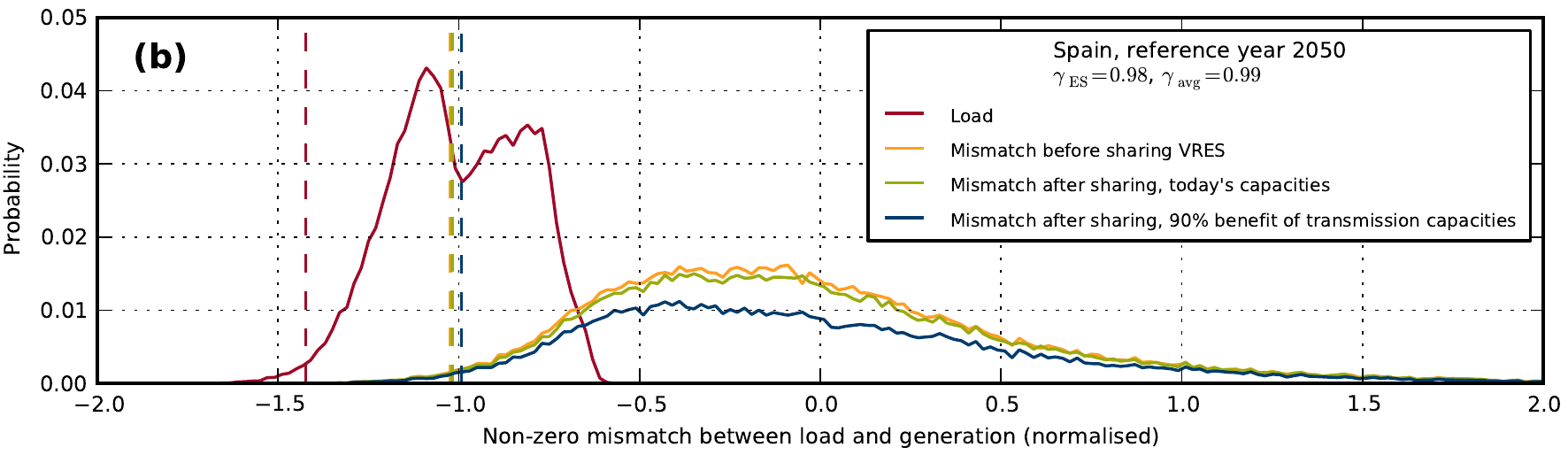}
    \caption{Distribution of the load (red), distribution of the mismatch
      between renewable generation and load before power sharing takes place
      (yellow), after sharing takes place with today's line capacities
      (green), and after sharing takes place with the 90\% benefit of
      transmission line capacities (blue), for the years 2030 (a) and 2050
      (b), for Spain. The dashed lines indicate the 99\% quantile of the
      residual deficit. For clarity, the peak at zero is not shown here.
      Some of the lines cover each other.
    }
    \label{fig:mmES}
  \end{center}
\end{figure}
\begin{figure}[!ht]
  \begin{center}
    \includegraphics[width=0.89\textwidth]{\figdir/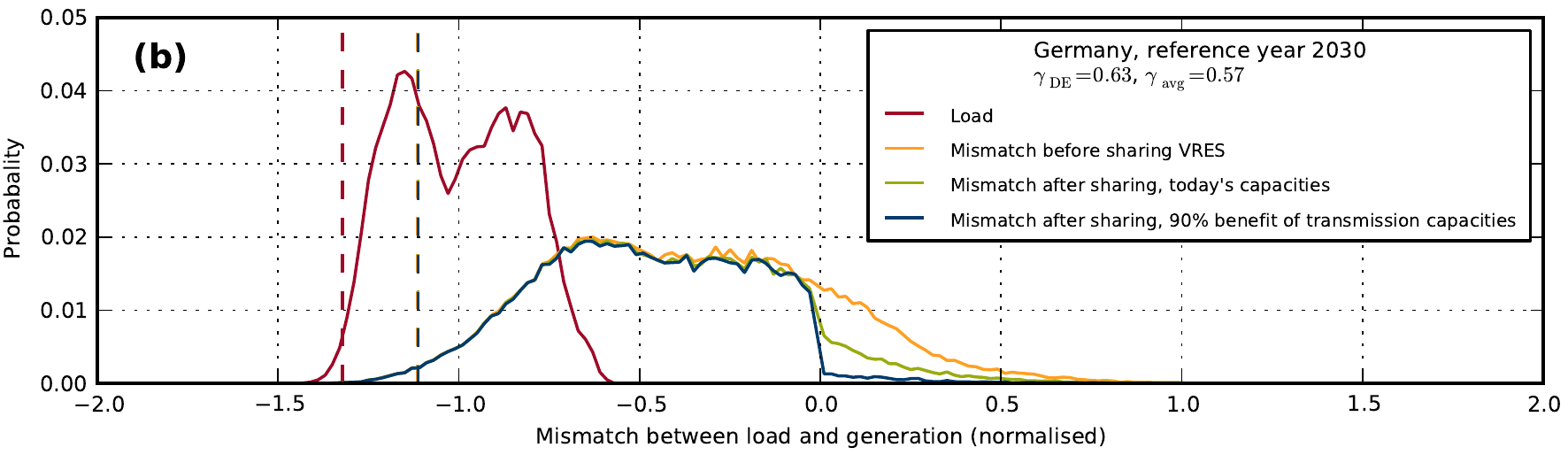}
    \includegraphics[width=0.89\textwidth]{\figdir/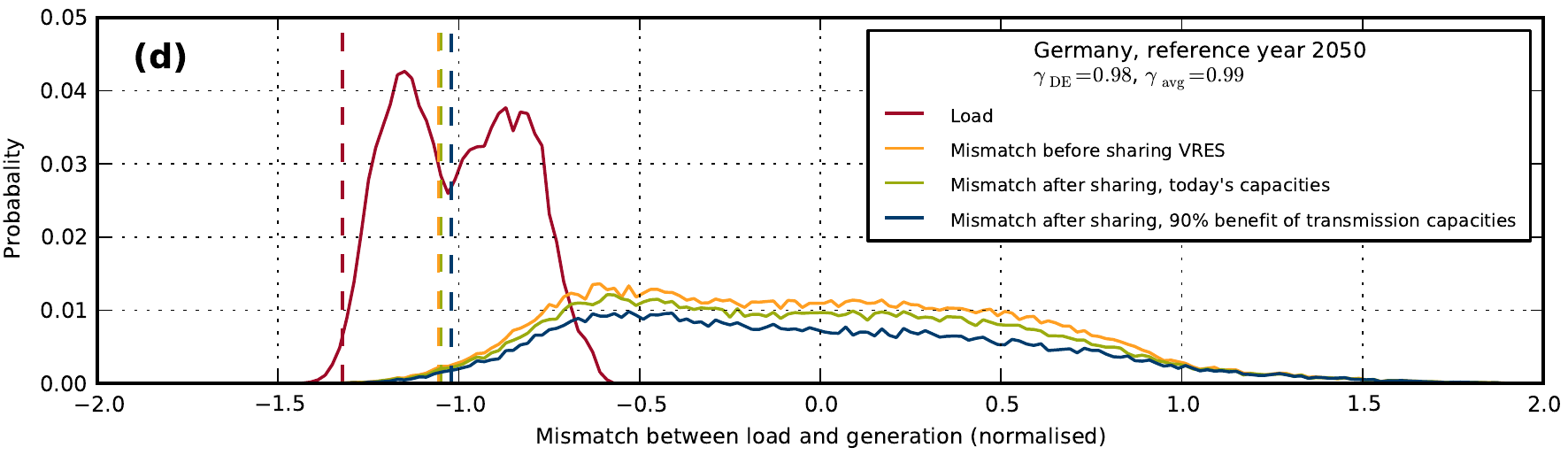}
    \caption{Distribution of the load (red), distribution of the mismatch
      between renewable generation and load before power sharing takes place
      (yellow), after sharing takes place with today's line capacities
      (green), and after sharing takes place with the 90\% benefit of
      transmission line capacities (blue), for the years 2030 (a) and 2050
      (b), for Germany. The dashed lines indicate the 99\% quantile of the
      residual deficit. For clarity, the peak at zero is not shown here.
      Some of the lines cover each other.
    }
    \label{fig:mmDE}
  \end{center}
\end{figure}

These findings are illustrated when looking at mismatch histograms, Figs.\
\ref{fig:mmDK}, \ref{fig:mmES}, and \ref{fig:mmDE}. For the influence of load
size, compare the situation in Germany and Denmark in 2050, \ref{fig:mmDK}b and
\ref{fig:mmDE}b. The mismatch before sharing renewables (yellow curves) is
comparable, since it is normalized by the mean load in both cases. After sharing
with the strong 90\% benefit of transmission capacity layout (blue curves),
Denmark's residual mismatch is much smaller than the German one. The correlation
between load size and fraction of the surplus that can be exported (Fig.\
\ref{fig:cor}b) shows that this is not just an accident, but a general trend.

The effect of position in the network is well illustrated when comparing Spain
(Fig.\ \ref{fig:mmES}b) and Denmark (Fig.\ \ref{fig:mmDK}b), and looking again
at mismatch before (yellow) and after sharing with the strong transmission
layout (blue). Again, we see that Spain's reduction is smaller than Denmark's.
The early export boom in Denmark and Spain can be guessed from the almost
complete elimination of the surplus production tail by exports in 2030 (Fig.\
\ref{fig:mmDK}a and \ref{fig:mmES}a). For more details and alternative
illustrations, see \cite{sarah}.

There are other general observations from the mismatch histograms, Fig.
\ref{fig:mmDK}-\ref{fig:mmDE}. One is that transmission is able to reduce the
bulk of the mismatch, roughly the region between $-1$ times the average load and
$+1$ times the average load. Here, enhanced transmission clearly leads to fewer
mismatch events. However, the tails of the distributions are almost unaffected.
This corresponds to a reduction of mismatch energy by transmission (which we
calculated above to be at most 40\% of the total mismatch energy in a fully
renewable scenario), but not so much of the "mismatch power capacity". From
these plots, it seems that large surplus as well as large deficit events hit
Europe more or less synchronously, thus preventing the countries from smoothing
the mismatch out by distributing it geographically. Ongoing research indicates
that this effect can be mitigated to some extend by sharing not only VRES
generation, but also balancing and surplus energy. A radical approach of
"complete sharing" is able to reduce the high quantiles of balancing from about
100\% of the average load to about 80\% of the average load, see Fig. 8 and the
corresponding discussion in \cite{rolando}.

The other observation to be made is that the transmission grid as it is is
highly inhomogeneous. While it does not make much of a difference for Denmark
whether we have the line capacities seen today or the 90\% benefit of
transmission layout, for Spain this is the crucial difference between being able
to participate in European trade or not (compare the green and the blue curves
in Fig.\ \ref{fig:mmDK}b and Fig.\ \ref{fig:mmES}b).


\section{Conclusions}

According to our analysis, transmission can reduce the balancing energy by up to
40\%, in particular by decreasing the bulk of the mismatch between generation
and load. On the other hand, transmission cannot take care of the remaining 60\%
balancing energy, and equivalently the corresponding 60\% surplus (for the fully
renewable end-point scenario). As for the power capacities of balancing and
alternative usage/curtailment, transmission (with a different flow paradigm than
presented here) is at best able to reduce them by about 20\% of the average
load.

If a reinforced transmission grid is to be build, we recommend to follow the
quantile capacity interpolation method. This enables Europe to gain 90\% of the
potential benefit of transmission while roughly quadrupling today's transmission
capacities. This build-up is comparable to what has been seen over the past
decade \cite{entsoe_ntc}.

\bibliographystyle{unsrt}
\bibliography{literatur}

\end{document}